\documentclass[10pt, a4paper, twocolumn, twoside]{article} 

\usepackage[english]{babel} 

\usepackage{microtype} 

\usepackage{amsmath,amsfonts,amsthm} 

\usepackage[svgnames]{xcolor} 

\usepackage[small, labelfont=bf, up]{caption} 

\usepackage{booktabs} 

\usepackage{lastpage} 

\usepackage{graphicx} 

\usepackage{enumitem} 
\setlist{noitemsep} 

\usepackage{sectsty} 
\allsectionsfont{\usefont{OT1}{phv}{b}{n}} 

\usepackage{geometry} 

\usepackage[T1]{fontenc} 
\usepackage[utf8]{inputenc} 

\usepackage{XCharter} 

\usepackage{journals}

\usepackage{fancyhdr} 
\newcommand{\shorttitle}[1]{\fancyhead[CE]{\textsl{#1}}}
\newcommand{\shortauthors}[1]{\fancyhead[CO]{\textsl{#1}}}

\fancypagestyle{firstpage}{ 
	\fancyhf{}
        \lhead{\vspace*{0.0em} \textit{\footnotesize 21$^{\rm st}$ European Workshop on White
            Dwarfs \\ Castanheira, Campos, Montgomery, eds. \\[-0.2em]
          July 23--27, 2018, Austin, Texas, USA}}
}

\date{}

\newcommand{\authorstyle}[1]{{\large\usefont{OT1}{phv}{b}{n}\color{DarkRed}#1}} 

\newcommand{\institution}[1]{{\footnotesize\usefont{OT1}{phv}{m}{sl}\color{Black}#1}} 

\usepackage{titling} 

\newcommand{\HorRule}{\color{DarkGoldenrod}\rule{\linewidth}{1pt}} 

\pretitle{
	\vspace{-30pt} 
	\HorRule\vspace{10pt} 
	\fontsize{16}{12}\usefont{OT1}{phv}{b}{n}\selectfont 
	\color{DarkRed} 
}

\posttitle{\par\vskip 10pt} 

\preauthor{} 

\postauthor{ 
	\vspace{0pt} 
	{\par\HorRule} 
	\vspace{-10pt} 
}

\newcommand{\newabstract}[1]{
    {\section*{Abstract}
    \bfseries #1}
  }

\usepackage[authoryear]{natbib}
\DeclareMathOperator{\sech}{sech}

\title{Constraining the Milky Way potential with Double White Dwarfs} 

\shorttitle{Constraining the Milky Way potential with Double White Dwarfs} 

\shortauthors{Korol, Rossi, and Barausse} 

\author{
        \authorstyle{V.~Korol,$^1$ E.~M.~Rossi,$^1$ and E.~Barausse$^2$}
	\newline\newline 
	$^1$\institution{Leiden Observatory, Leiden University, PO Box 9513, 2300 RA, Leiden, the Netherlands; 
          korol@strw.leidenuniv.nl}\\ 
	$^2$\institution{Institut d'Astrophysique de Paris, CNRS \& Sorbonne Universit{\'e}s, UMR 7095, 98 bis Bd Arago, 75014 Paris, France} 
      }

\begin{document}

\maketitle 

\thispagestyle{firstpage} 


\newabstract{
  The upcoming LISA mission is the only experiment that will allow us to study the Milky Way's structure using gravitational wave signals from Galactic double white dwarfs (DWDs).
  The total number of expected detections exceeds $10^5$.
  Furthermore, up to a hundred DWDs can be simultaneously detected in both gravitational and optical radiation (e.g. with Gaia and LSST as eclipsing),
  making DWDs ideal sources for performing a multi-messenger tomography of the Galaxy.
  We show that LISA will detect DWDs everywhere, mapping also the opposite side of the Galaxy. This complete coverage will : 
  (1) provide precise and unbiased constraints on the scale radii of the Milky Way's bulge and disc, and 
  (2) allow us to compute the rotation curve and derive competitive estimates for the bulge and disc masses, when combining gravitational wave and optical observations.
  }


\section{Introduction}

So far we have only explored the Milky Way through electromagnetic (EM) radiation, but the upcoming LISA mission will allow us to probe our Galaxy with gravitational wave (GWs) radiation.
Moreover, by exploiting GWs we can circumvent some difficulties inherent to EM observations, such as the dust extinction at optical wavelengths, luminosity bias and various difficulties related to distance determination.
LISA is a space-based GW experiment approved by ESA in 2017 and scheduled for launch in early 2030 \citep{ama17}.
Its design consists of a set of three identical spacecrafts in an equilateral triangle formation, with a center of mass that will follow a circular heliocentric trajectory lagging behind the Earth.
The spacecraft separation is set to $2.5 \times 10^6 \,$km, making LISA sensitive to GW frequencies between $0.1$ and $100\,$mHz.
Our Galaxy hosts a variety of stellar mass mHz GW sources like detached DWD binaries, AM CVn systems, ultra-compact X-ray binaries and binary black holes. 
A number of these binaries have been identified though EM radiation \citep[e.g.,][]{kup18}.
In particular, detached DWDs are foreseen to be the most abundant LISA sources, with $>10^5$ individually resolvable in frequency \citep[e.g.,][]{nel04,rui10,mar11,kor17}.
Indeed, these binaries are predicted to be so common in the Milky Way that overlapping signals of those below the LISA sensitivity threshold will form a background noise for the mission \citep[e.g.,][]{rob17}. 
Both resolved DWDs and the DWD background bear the imprint of the overall Galactic stellar population, thus both can be used to study the Milky Way's structure and star formation history.
By fitting density distributions of resolved LISA detections we can precisely estimate the scale radii of the disc and the bulge, while by fitting the DWD background we can recover the disc scale height \citep{ben06,ada12,kor18}.
Additional information (such as the motion of DWDs) required to constrain the Milky Way potential can be recovered from optical observations. 
By fitting the Galactic rotation curve constructed using distances inferred from gravitational waves and proper motions from optical observations one can obtain competitive estimates of the bulge and the disc masses \citep{kor18}.


\section{Synthetic Milky Way}

To obtain a mock DWD population we exploit the population synthesis code SeBa \citep{por96,too12}.
The initial stellar population is obtained assuming the Kroupa initial mass function, a flat binary mass ratio distribution, a log-flat distribution for the binary semi-major axis, 
a thermal distribution for the orbit eccentricity, an isotropic distribution for binary inclination angles, and a constant binary fraction of 50\% \citep[][]{kro93,duc13,rag10}.
We use the $\gamma \alpha$ prescription for the common envelope phase \citep{nel01}. 
The sensitivity of our model to these assumptions is discussed in \citet{kor17} and \citet{too17}.
We adopt a potential composed of a bulge, a stellar disc and a dark matter (DM) halo.
The density distribution of the disc component for our model can be analytically expressed as
\begin{equation}  \label{eqn:disc}
\rho_{\rm disc}(t,R,z) = \rho_{\rm BP}(t)\, e^{-R/R_{\rm d}}\sech^2 \left( \frac{z}{Z_{\rm d}} \right) \ {\text M}_{\odot}\, {\text{kpc}}^{-3},
\end{equation} 
where $0 \le R \le 19$ kpc is the cylindrical radius measured from the Galactic center, $R_{\rm d} = 2.5\,$kpc is the characteristic scale radius, 
$Z$ is the height above the Galactic plane and $Z_{\rm d}=300\,$pc is the characteristic scale height of the disc. 
To account for the star formation history  we use the plane-projected star formation rate from \citet{boi99}, $\rho_{\rm BP}$, assuming the age of the Galaxy to be 13.5 Gyr.
The total mass of the disc in our model is $5 \times 10^{10}\,$M$\odot$ and the distance of the Sun from the Galactic center is $R_{\odot}=8.5\,$kpc.
The bulge component is modeled by doubling the star formation rate in the inner $3\,$kpc of the Galaxy and distributing DWDs spherically: 
\begin{equation} \label{eqn:bulge}
\rho_{\rm bulge}(r) = \frac{M_{\rm b}}{(\sqrt{2\pi} r_{\rm b})^3} e^{-r^2/2 r_{\rm b}^2}\  {\text M}_{\odot}\, {\text{kpc}}^{-3},
\end{equation}
where $r$ is the spherical distance from the Galactic center, $M_{\rm b} =  2.6 \times 10^{10}\,$M$\odot$ is the bulge total mass, and $r_{\rm b}=0.5\,$kpc is the characteristic radius of the bulge.
Finally, to describe the DM halo we use the \citet{nav96} density profile:
\begin{equation} \label{eqn:halo}
\rho_{\rm DM} (r) = \frac{\rho_{\rm h}}{(r/r_{\rm s})(1+r/r_{\rm s})^2} \ {\text M}_{\odot}\, {\text{kpc}}^{-3},
\end{equation}
where $r_{\rm s} = 20\,$kpc is the scale length of the halo and  $\rho_{\rm h} = 0.5 \times 10^7\,$M$_{\odot}$kpc$^{-3}$ is the halo scale density.
The total mass of the halo enclosed in a $100\,$kpc radius is equal to $4.8 \times 10^{11}\,$M$_{\odot}$.
We summarize the values of the  parameters adopted for our Milky Way fiducial model in Table \ref{tab:1}.


\section{Milky Way map in gravitational waves}

\begin{figure*}[t]
  \centerline{\includegraphics[width=2.0\columnwidth]{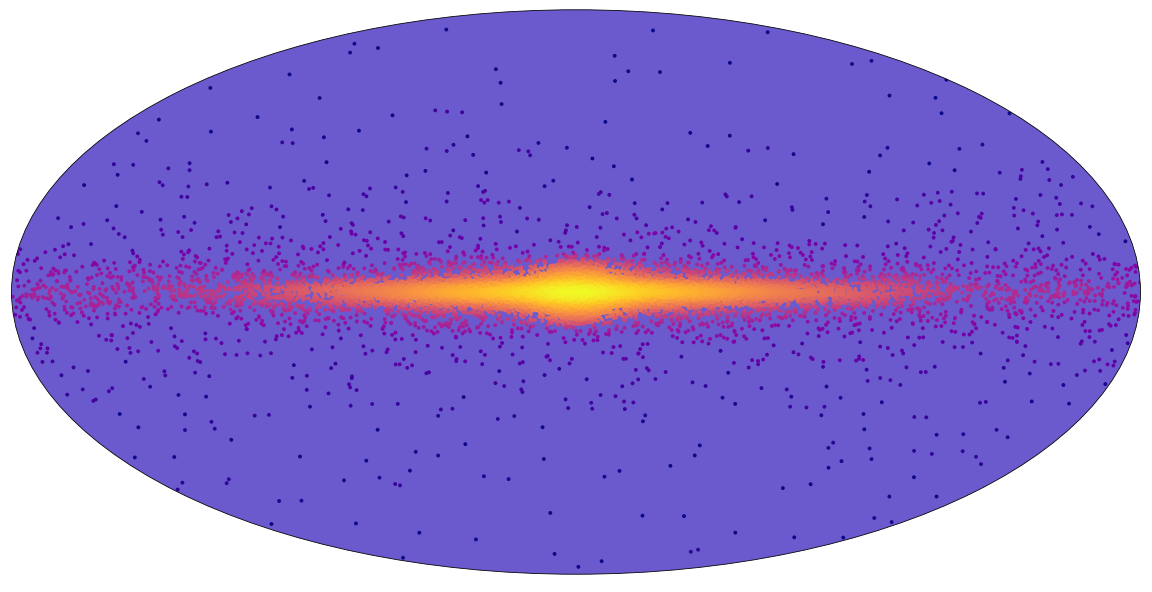}}
  \caption{Distribution of DWD detected by LISA in Galactic coordinates $(l,b)$. The color represents the density of DWDs: yellow being the most dense.} 
  \label{fig:1}
\end{figure*}

To model GW signals of DWDs we employ the Mock LISA Data Challenge (MLDC) pipeline designed for the simulation and analysis of a large number of quasi-monochromatic GW sources \citep[e.g.,][]{lit11}. 
The MLDC pipeline describes GW signals in terms of nine GW parameters: amplitude $A$, frequency $f=2/P$ with $P$ being the binary orbital period, $\dot{f}, \ddot{f}$, sky location in the LISA reference frame ($\theta$, and $\phi$), 
orbital inclination $\iota$,  GW polarization $\psi$ and the initial phase $\phi_0$.
For a given synthetic instrument noise curve, observation time and detection threshold, the pipeline provides a catalog of sources with signal-to-noise ratio (SNR) above the detection threshold 
and the uncertainties of the nine GW parameters.
We adopt a mission duration of $4\,$yr and the noise curve from \citet{ama17}, which corresponds to the mission design approved by ESA.
Finally, we set the SNR detection threshold to 7.

We find $2.6 \times 10^4$ DWDs in our mock Milky Way above the threshold. In particular, we estimate their distances and sky localization with their respective uncertainties.
We compute the distances from GW observables $A, f$ and $\dot{f}$ as \citep{mag08}
\begin{equation}
d = \frac{5 c}{96\pi^2}\frac{\dot{f}}{f^3 A},
\end{equation}
and the respective errors as
\begin{equation} \label{eqn:sigmad}
\frac{\sigma_{\rm d}}{d} \simeq  \left[\left(\frac{\sigma_{A}}{A}\right)^2 + \left(\frac{3\sigma_f}{f}\right)^2 + \left(\frac{\sigma_{\dot{f}}}{\dot{f}}\right)^2 \right]^{1/2},
\end{equation}
where ${\sigma_A}/{A}, {\sigma_f}/{f}$ and ${\sigma_{\dot{f}}}/{\dot{f}}$ are outputs of the MLDC pipeline.
We verify that the terms containing correlation coefficients are at most of the order of 1\%, and we thus neglect them in eqn.~\eqref{eqn:sigmad}.
The coordinate transformation between the LISA reference frame $(\theta,\phi)$ and the Galactic coordinate frame $(l,b)$ is outlined in \citet{kor18}.
The distribution of LISA detections in Galactic coordinates is shown in Fig.~\ref{fig:1}.
The map reveals a spherical agglomeration of stars in the center, which represents the bulge, and an extended flat distribution of stars, which represents the stellar disc.
We find that LISA will be able to detect these binaries to large distances, comparable with the extension of the disc, also mapping of the opposite side of the Milky Way \citep[see][figure 3]{kor18}.
Out of $2.6 \times 10^4$ LISA detections in our simulation, almost $8\times 10^3$ DWDs ($30\%$) have relative distance errors $< 30\%$.
In particular, a subsample of $\sim 100$ DWDs ($0.4\%$ of the catalog) has relative errors on the distance $< 1\%$. 
These are  high frequency ($>3\,$mHz), high SNR ($>100$) sources spread around the Galaxy. 
This remarkable precision is due to the fact that GW amplitudes decrease more slowly with distance ($\propto 1/d$) than EM observations ($\propto 1/d^2$). 
This property of GWs is at the heart of the unique ability of the LISA mission to study the Milky Way's structure. 
Finally, we find that on average LISA will locate DWDs within $1 - 10\,$deg$^2$ on the sky.

\subsection{Density distributions} \label{sec:31}

\begin{figure}[]
        \centering
	 \includegraphics[width=0.5\textwidth]{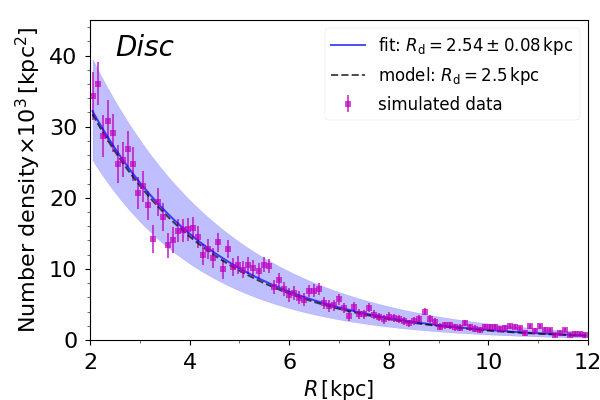} 
	 \includegraphics[width=0.5\textwidth]{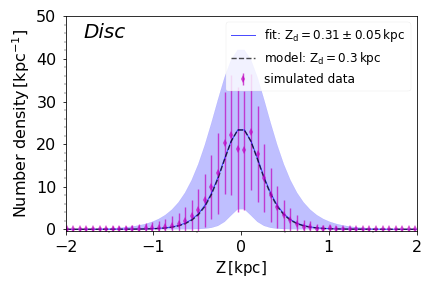}
	 \includegraphics[width=0.5\textwidth]{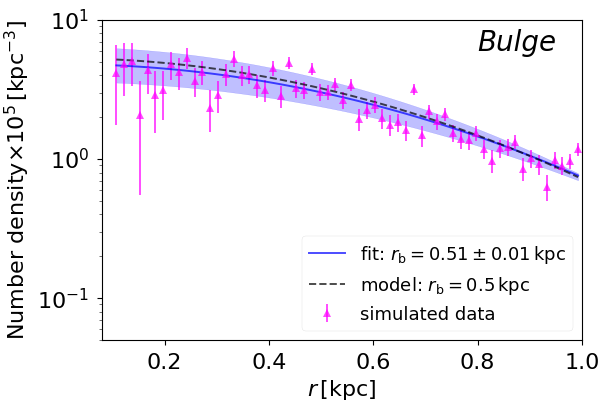}
         \caption{Number density profiles for LISA detections as a function of cylindrical radius $R$ (top panel), height above the Galactic plane $Z$ (middle panel), 
         and spherical radius $r$ from the Galactic center (bottom panel). Magenta points represent simulated data. 
         The blue solid line and the shaded area shows the best fit model and its $3\sigma$ uncertainty region. The dashed gray line shows the true number density.}
       \label{fig:2}
\end{figure}

Figure \ref{fig:1} suggests that LISA has the potential to reconstruct the density profiles of both the disc and bulge components of our mock Galaxy.
To test this we select binaries obtained from the LISA catalog with distances determined to better than 30\%.
First, we derive the radial density profile. We compute $10^5$ realizations of the 3D binary positions in the Galaxy by randomly drawing $l,b$ and $d$ 
from Gaussian distributions centered on their true values and with standard deviations provided by the MLDC pipeline. 
For each realization we compute the cylindrical Galactocentric distance, $R$, and we select those with $2 \le R \le 12\,$kpc. 
We bin DWDs in cylindrical shells of width $dR = 125\,$pc, and divide the bin counts by the shell volume.
We compute the error on the number density in each bin as the standard deviation over different realizations.
The obtained disc radial density profile is shown in the upper panel of Fig. \ref{fig:2}. 
Then, we fit the density profile with the analytic expression in eqn.~\eqref{eqn:disc}, obtaining the value for the disc scale radius of $R_{\rm d} = 2.54 \pm 0.08\,$kpc, in agreement with the fiducial value.
The blue solid curve shows the best fit, and the colored area shows the $3\sigma$ interval.

Next, we study the vertical distribution of DWDs in the disc.
Again, we select only  binaries  with  $2 \le R \le 12\,$kpc, and we bin them in the $R-Z$ plane.
In each radial bin, we model the number density with a $\sech^2 (Z/Z_{\rm d})$ function and fit $Z_{\rm d}$ to test whether the scale height is constant with $R$.
We find a constant behavior and therefore we decide to increase the statistics by computing the average value of $Z_{\rm d}$ and its error by using a stacked radial profile. 
In this way, we find $Z_{\rm d} = 0.31 \pm 0.04\,$kpc, which is consistent with the fiducial value  of $0.3\,$kpc.

Finally, to estimate the scale radius of the bulge we select DWDs in the inner $1\,$kpc, and we compute $10^5$ realization of the binary positions in the Galaxy, as we did for the disc radial profile.
For each realization we  estimate the number density profile by counting DWDs in spherical shells with radius $r$ and $dr = 20\,$pc, dividing this number by the shell volume.
The result is given by the magenta triangles in the bottom panel of Fig. \ref{fig:2}.
To fit the scale radius of the bulge, we use eq.~\eqref{eqn:bulge} as the model distribution, and we obtain $r_{\rm b}=0.51 \pm 0.013\,$kpc.

\section{Combining gravitational wave and optical observations}

The mass of the Milky Way's disc and bulge cannot be derived from GW data alone.
Optical observations of the motion of DWDs in the sky are required to constrain them.  
The sky localization from GW data is typically poor, which  makes it difficult to identify counterparts to GW sources in EM databases. 
In practice, to assemble a sample of optical counterparts, one possibility is to search in optical surveys for  periodically variable sources with a frequency and within an area on the sky matching those provided by LISA. 
To assess whether this is possible we focus on edge-on binaries, which allow for better distance estimation with GWs and are also easy to identify by optical telescopes as eclipsing sources \citep{shah12}. 
In our previous study we found that {\sl Gaia} \citep{gaia16} and the Large Synoptic Survey Telescope \citep[LSST,][]{lsst09} can identify $75$ of DWDs detected by LISA as eclipsing (of which 23 are identified by both) \citep{kor17}.
We use this sample of binaries also in this work.
To understand how well it will possible to constraint the motion of the DWD optical counterparts we simulate their parallaxes $\varpi$ and proper motions $\mu$ as measured by {\sl Gaia} and LSST.
We draw the parallaxes $\varpi$ from a Gaussian distribution centered on $1/d$ and with standard deviation $\sigma_{\varpi}$.
We compute proper motions as $\mu = V_{\rm t}/d$, where $V_{\rm t}$ is the tangential velocity, which for a fixed Galactic potential can be derived from geometrical consideration as outlined in \citet{kor18}.
The end-of-mission parallax and proper motion errors ($\sigma_{\varpi}$ and $\sigma_{\mu}$) for {\sl Gaia} can be estimated with the {\sl pyGaia}\footnote{https://github.com/agabrown/PyGaia} python toolkit, and for LSST they are tabulated in \citet{lsst09}.
For those DWDs that are identified by both optical instruments we select the measurement with the smallest uncertainty.
Next, we convert parallaxes into distances following the Bayesian inference approach indicated in \citet{bai15,lur18}, adopting an exponentially decreasing space density function with scale length of $400\,$pc as the prior. 
Our choice for the scale length is explained in \citep{kup18}.
We find that the expected relative error in parallax for binaries at $d<1\,$kpc is $<20\%$, however this is only $30\%$ of the optical counterparts. 
This gives us the motivation to combine distances derived from parallaxes with the LISA measurements.
Again, this can be done by using Bayes' theorem and is described in \citet{kor18}, Sect.~3.3.
The resulting distances (hereafter $d_{\rm obs}$) represent the harmonic means of the two measurements (GW and optical) and the respective errors are equal to twice the harmonic mean of the individual errors.
Essentially, in this way we select the best between GW and optical measurements and we reduce the uncertainty on the distance compared to just selecting the more precise of the EM or GW measurements individually.

\subsection{Rotation curve}

\begin{table} 
\begin{center}
\caption{Scale parameters of the adopted Milky Way model.}
\label{tab:1}
\centering
 \begin{tabular}{| c | c | c| c|}
 \hline
   Parameter & Fiducial & Reconstructed\\ \hline \hline
   $M_{\rm b}\times 10^{10}\,$M$_{\odot}$ & $2.6$ & $2.49^{+0.44}_{-0.42}$ \\ 
   $r_{\rm b}$ kpc & $0.5$ & $0.51 \pm 0.01$\\ \hline
   $M_{\rm d}\times 10^{10}\,$M$_{\odot}$ & $5$  & $5.30^{+1.29}_{-1.71}$\\ 
   $R_{\rm d}$ kpc & $2.5$ & $2.54 \pm 0.08$ \\
   $Z_{\rm d}$ kpc & $0.3$ & $0.31 \pm 0.05$\\  \hline
   $\rho_h\times 10^7\,$M$_{\odot}$/kpc$^{3}$ &  $0.5$ & $0.67^{+0.77}_{-0.38}$ \\
   $r_{\rm h}$ kpc & $20$  & $15.19^{+7.50}_{-4.02}$\\ \hline
  \end{tabular}
\end{center}
\end{table}

\begin{figure}
        \centering
	 \includegraphics[width=0.5\textwidth]{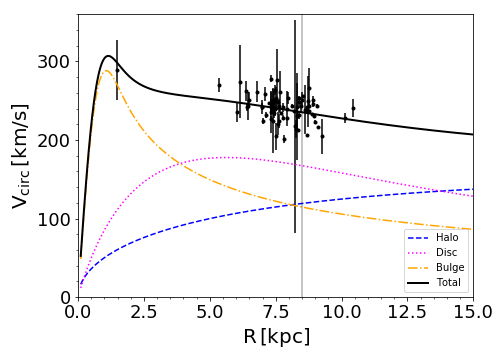} 
         \caption{Rotation speed of DWDs with EM counterparts. 
         The black solid curve shows the model's rotation curve. 
         Colored lines represent the contributions of different Galactic components to the total rotation curve.
         The vertical line marks the position of the Sun.}
       \label{fig:3}
\end{figure}

\begin{figure*}[t]
        \centering
	 \includegraphics[width=1.7\columnwidth]{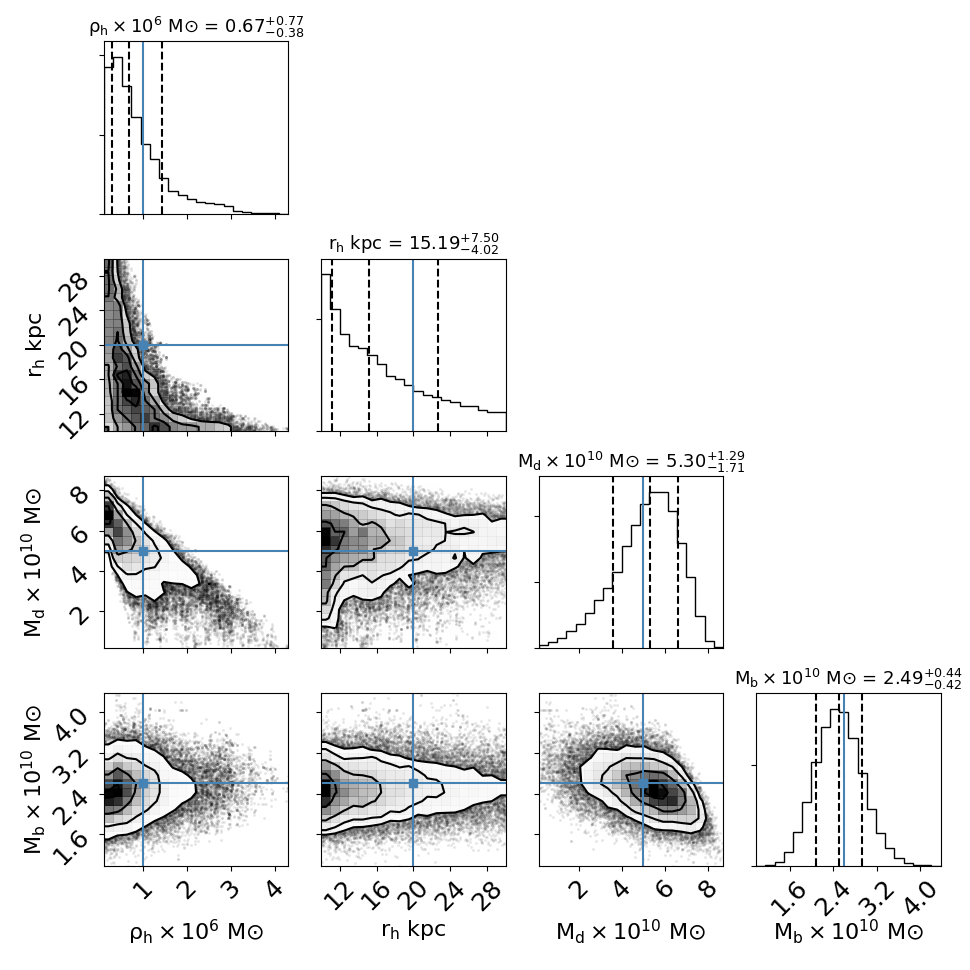} 
         \caption{The posterior probability distribution of the four free parameters of our rotation curve fitting model, $M_{\rm b},  M_{\rm d}, \rho_h$ and $r_{\rm h}$. 
         Blue lines mark the true values listed in Tab. \ref{tab:1}}
       \label{fig:4}
\end{figure*}

To simulate the motion of DWD in the Galaxy we compute the observed rotation speed (in km/s) as
\begin{equation} \label{eqn:vrot_obs}
V_{\rm obs}(R) = - \frac{R}{d_{\rm obs} - R_0 \cos l} \left(4.74 \mu_{\rm obs} d_{\rm obs} + V_0 \cos l \right),
\end{equation}
where $d_{\rm obs}$ and $\mu_{\rm obs}$ are the observed distance and proper motion, $R_0 = 8.5\,$kpc and $V_0 = 235\,$km/s are the position and the rotation velocity of the Sun.
To simulate {\sl Gaia} and LSST measurements of DWD proper motions, we assign an observed proper motion $\mu_{\rm obs}$ to a source by sampling from a Gaussian centered on $\mu$ and with an error $\sigma_{\mu}$.
For each DWD, we calculate $V_{\rm obs}(R)$ for $10^5$ independent realizations of $\mu_{\rm obs}$ and $d_{\rm obs}$, 
and we assign an observed velocity and a measurement error equal, respectively, to the mean and the standard deviation of the resulting distribution of $V_{\rm obs}(R)$.  
The result is represented in Fig. \ref{fig:3}.
We fit the obtained data with the total rotation curve corresponding to our Milky Way model, 
which we compute as $V^2(R) = R d\Phi/dR$ with $\Phi$ being the total potential.
We perform the fit with an MCMC code fixing $r_{\rm b}, R_{\rm d}$ and $Z_{\rm d}$ to the values obtained in Sect.\ref{sec:31}, 
and leaving $M_{\rm d}$ and $M_{\rm b}$, $\rho_{\rm h}$, and $r_{\rm h}$ free, but confined by flat uninformative priors in the respective ranges: 
$(1 - 10) \times 10^{10}$M$_{\odot}$, $(0.1 - 10) \times 10^7\,$M$_{\odot}$/kpc$^3$ and $10 - 30\,$kpc.
The final posterior probability distribution for the free parameters is represented in Fig. \ref{fig:4}.
It reveals that we can recover the mass of the disc and bulge components, but not that of the DM halo.
This is because there is no optical data at $R > 11\,$kpc, where the halo component dominates the dynamics in our Milky Way model (Fig.\ref{fig:3}).
We estimate the mass of the disc to be $M_{\rm d} = 5.3^{+1.29}_{-1.71} \times 10^{10}\,$M$_{\odot}$ and the mass of the bulge to be $M_{\rm b} = 2.49^{+0.44}_{-0.42} \times 10^{10}\,$M$_{\odot}$, in good agreement with our fiducial values. 
Remarkably, our constraints on the bulge mass are extremely competitive with those derived from EM tracers \citep[see e.g.][for a review]{bla16}. 
The larger errors on the disc mass stem from our choice to leave the halo parameters unconstrained.
Thus, an improvement of this analysis should involve the adoption of stringent constraints on the DM halo parameters.
We summarize all the recovered parameters in Tab.~\ref{tab:1}.

\vspace*{-0.5em}
\section{Conclusions}

In this study we introduce the idea of the multi-messenger study of the Milky Way using Galactic DWD binaries, 
and we investigate the prospects for tracing the baryonic mass of the Milky Way using GWs in combination with their optical counterparts. 
The advantages over traditional tracers include the possibility of looking through the bulge, and beyond, thus allowing one to map both sides of the Galaxy using the same tracer.
We show that the unique property of DWDs allows one to recover the scale radii of the baryonic components with a percent precision. 
In synergy with optical data, GW measurements will provide competitive mass estimates for the bulge and stellar disc.
However, our choice to use GW sources and their EM counterparts limits our ability to  constrain the DM halo component of the Galaxy. 
This highlights the importance of a more precise knowledge of the DM halo to improve baryonic mass measurements. 

\section*{Acknowledements}
We acknowledge the Gaia Project Scientist Support Team and the Gaia Data Processing and Analysis Consortium (DPAC) for developing {\sl pyGaia} python tool kit that we used to simulate {\sl Gaia} data.
This project has received funding from the European Union’s Horizon 2020 research and innovation programme under the Marie Sklodowska-Curie grant agreement No 690904.
EMR, VK acknowledge NWO WARP Program, grant NWO 648.003004 APP-GW.



\end{document}